*International Journal of Computing & Information Sciences*  Vol. 4, No. 3, December 2006   97
A Decision Support Toll for Assessing the Maturity of the Software Product Line Process
Faheem Ahmed and Luiz Fernando Capretz
Pages: 97-115

# A Decision Support Tool for Assessing the Maturity of the Software Product Line Process

Faheem Ahmed         Luiz Fernando Capretz

**Department of Electrical and Computer Engineering**
**University of Western Ontario, London, Ontario, N5A5B9, Canada**
fahmed@engga.uwo.ca         lcapretz@eng.uwo.ca

**Abstract:** *The software product line aims at the effective utilization of software assets, reducing the time required to deliver a product, improving the quality, and decreasing the cost of software products. Organizations trying to incorporate this concept require an approach to assess the current maturity level of the software product line process in order to make management decisions. A decision support tool for assessing the maturity of the software product line process is developed to implement the fuzzy logic approach, which handles the imprecise and uncertain nature of software process variables. The proposed tool can be used to assess the process maturity level of a software product line. Such knowledge will enable an organization to make crucial management decisions. Four case studies were conducted to validate the tool, and the results of the studies show that the software product line decision support tool provides a direct mechanism to evaluate the current software product line process maturity level within an organization.*

Received: September 30, 2005 | Revised: July 10, 2006 | Accepted: August, 12, 2006

## 1. Introduction

The software product line has emerged as an appealing phenomenon within organizations dealing with software development. The concept of the software product line, proposed by the Software Engineering Institute (SEI), is a comprehensive model for an organization building applications based on common architecture and core assets [12]. The concept of the software product line is based on the development of identical systems having controlled variability among each other. Clements and Northrop [4] define the software product line as a set of software-intensive systems sharing a common, managed set of features that satisfy the specific needs of a particular market segment or mission and are developed from a common set of core assets in a prescribed way. The software product line deals with the assembly of products from current core assets, commonly known as components, and the continuous growth of the core assets as production proceeds. This idea has become vital in terms of software development from component-based architecture. The overall engineering efforts during software product line development can be divided into the following three essential interrelated activities:

**Core Asset Development:** Core assets in a software product line may include architecture, reusable software components, domain models, requirement statements, documentation, schedules, budgets, test plans, test cases, process descriptions, modeling diagrams, and other relevant items used for product development. There is no specific definition for core asset inclusion, except that it is an entity used for development purposes. The goal of core asset development is to establish the production capability of developing products [11]. The major inputs to the core asset development activity are: product constraints, styles, patterns, frameworks, production constraints, production strategy, and the inventory of pre-existing assets. The outputs of core assets development are software product line scope, core assets and the production plan. Software product line scope describes the characteristics of the products developed. The production plan gives an in-depth picture how products will be developed from core assets. Core assets are those entities that may be used in the product development. The collection of core assets is termed as core asset repository and the initial state of the core asset repository depends upon the type of approach being used to adopt software product line approach within an organization.

**Product Development:** In product development activity, products are physically developed from the core assets, based on the production plan, in order to satisfy the requirements of the software product line. The essential inputs of product development activity are requirements, product line scope, core assets and the production plan. Requirements describe the purpose of the product line along with functionalities and characteristics of the products developed. Product line scope describes qualification criteria for a product to be included or excluded from software product line based on functional and non-functional characteristics. The production plan describes a strategy to use the core assets to assemble products. A product line can produce any number of products depending upon the scope and requirements of the software product line. The product development activity iteratively communicates with core asset activity and adds new core assets as products are produced and software product line progresses.

**Management:** Management plays a vital role in successfully institutionalising the software product line within an organization, because it provides and coordinates the required infrastructure. Management activity involves essential processes carried out at technical and organizational levels to support the software product line process. It ensures that necessary resources must be available and well coordinated. The objective of "Technical Management" is to oversee the core asset and product development activities by ensuring that the groups who build core assets and the groups who build products are engaged in the required activities, and are following the processes defined for the product line [4]. Technical management plays a critical role in decision-making about the scope of software product line based on requirements. It handles the associated processes of software development. Northrop [11] summarized the responsibilities of organizational management, which are: structuring an organization, resource management and scheduling, cost control and communication. Organizational management deals in providing a funding model for the software product line in order to handle cost constraints associated with the project. It ensures a viable and accurate communication and operational path between essential activities of software product line development because the overall process is highly iterative in nature. The fundamental goal of the organizational management is to establish an adoption plan, which completely describes a strategy to achieve the goals of software product line within an organization. The major responsibility of the management is to ensure proper training of the people to become familiar with the software product line concepts and principles. Management deals with external interfaces for smooth and successful product line and performs market analysis for internal and external factors to determine the success factor of software product line. Management performs organizational and technical risk analysis and continues tracking critical risk throughout the software product line development.

## 1.1 Problem Definition

The maturity of the software product line process is a growing concern within organizations. Management requires a certain methodology and a particular tool in order to evaluate the process maturity. This evaluation certainly helps them in making management decisions in technical and organizational domains. As previously discussed, the software product line process is composed of three activities: core asset development, product development and management. In order to facilitate management decisions, it is essential to know how well these activities are being performed in an ongoing project. Research [2, 6, 8, 9] has been conducted on the process definition and on the associated relevant activities of the software product line. Although the software product line is gradually gaining popularity over due to its economical impact [3], there has not been a great deal of research in establishing an appropriate approach for software product line process assessment. Thus, the aim of this paper is to provide structural and architectural implementation and application details of a decision support tool based on the framework [1], which provides a methodology to evaluate the process maturity of the software product line. This tool is named Software Product Line Decision Support Tool (SPLDST). The proposed tool is based on fuzzy logic, an approach chosen to handle the uncertainty and imprecision of the process input. To support and validate the implementation of the tool, a number of experiments were conducted by using actual industrial data drawn from reputable, well-known software organizations.

## 1.2 Software Product Line Process Assessment: Related Work

Jones and Soule [7] discussed the relationship between the software product line process and the CMMI model, observing that the software engineering process discipline as specified in CMMI models provides an important foundation for software product line practice. These researchers concluded that the software product line requires mastery of many other essential practice areas apart from the key process areas of the CMMI model. Although they have compared the process areas of the software product line and the CMMI and found some similarities, there is still a need to establish a comprehensive strategy for process assessment particularly for the software product line. The Software Engineering Institutes (SEI) proposed Product Line Technical Probe (PLTP), which is aimed at discovering an organization's ability to adopt and succeed with the software product line approach. The PLTP is based on the framework for software product line practice [4]. In this framework, there are 29 practice areas, which are classified into the three categories of product development, core asset development and management. The framework does not specifically assign levels to organizations based on the maturity of the current process; rather, it simply identifies those potential areas of concern that require attention during software product line activity.

Linden et al. [10] proposed a four-dimensional software product family engineering evaluation model based on the configuration of BAPO, in which the dimensions are composed of business, architecture, process, and organization. The proposed framework is an outcome of the European projects Concepts to Application in System-Family Engineering (CAFÉ) and Engineering Software Architecture, Process and Platforms for System Family Engineering (ESAPS). Each dimension identifies an evaluation scale, and the overall evaluation of an organization will create a profile with separate values for each of the four scales. The Business Evaluation is divided into a scale of five levels: Reactive, Awareness, Extrapolate, Proactive and Strategic. The Architecture Dimension also has five levels: Independent Product Development, Standardized Infrastructure, Software Platform, Software Product Family and Configurable Product Base. Similarly, the Process Dimension is divided into the categories of Initial, Managed, Defined, Quantitatively Managed and Optimizing. Finally, the organizational evaluation contains is classified into Unit Oriented, Business Lines Oriented, Business Group/Division, Inter Division/Companies and Open Business. A more comprehensive strategy for the maturity assessment of the software product line will be released as an outcome of a recently initiated European project named FAMILIES [5].

Ahmed and Capretz [1] proposed rules for developing and managing a software product line within an organization. On the basis of the proposed rules, a fuzzy logic-based software product line process assessment framework was proposed. In order to evaluate the reliability of the proposed framework, Ahmed and Capretz compared the results of the proposed software product line process assessment approach with the exiting CMMI levels achieved by the organizations under study. The fuzzy logic approach presented in that work transforms the software product line process variables into CMMI levels as output. The purpose of this transformation was to investigate the extent of reliability of the proposed approach with an existing standardized approach like CMMI. Another aspect of CMMI involvement with that framework was to investigate the impact of already achieved CMMI level on software product line process. The case studies presented in that work was used to find out how effectively an organization can execute software product line process when it has already achieved a higher CMMI level. As a result, this paper addresses the architectural structure and modeling descriptions of the software product line decision support tool of that framework.

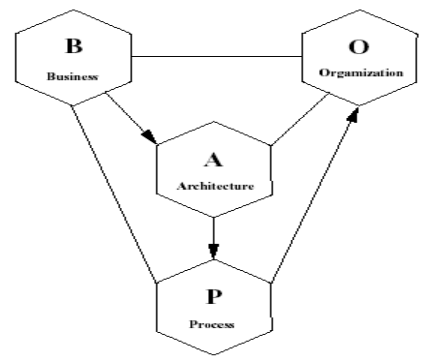

**Figure 1: BAPO model of software product family**

## 2. Fuzzy Logic System

The term "fuzzy logic", introduced by Zadeh [13], is used to handle situations where precise answers cannot be determined. Fuzzy logic is a form of algebra, which deals with a range of values from "true" to "false" for the purpose of decision-making with imprecise data. Zadeh [14] explained that the purpose of fuzzy logic is to provide a variety of concepts and techniques for representing and inferring from knowledge that is imprecise, uncertain or lacking reliability. The fuzzy logic inference system involves various steps to process the input and to produce output. These steps will be discussed below:

**Step-0 Linguistic Variable and Membership Mapping:** Linguistic variables take on linguistic values in fuzzy logic in the same way that numerical variables have numerical values. Linguistic variables are words commonly known as linguistic; for example, in order to describe height, there may be three linguistic variables: short, average, and tall. Each linguistic term is associated with a fuzzy set, each of which has a defined membership function (MF). A membership function is a curve that defines the way in which each point in the input space is mapped to a membership value between 0 and 1. For example, one can consider a universal range of 40 inches to 90 inches for the height of a person as well as three linguistic variables such as short, average, and tall. Figure 2 shows the mapping between linguistic variables and fuzzy membership.

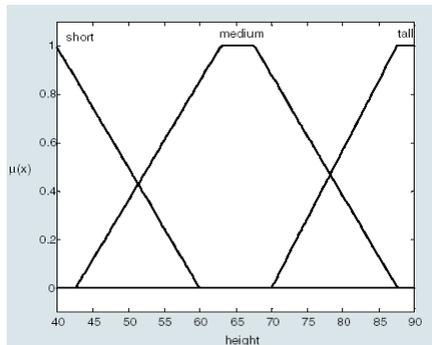

**Figure 2: Linguistic variable and fuzzy membership mapping**

**Step-1 Fuzzification:** Fuzzification is the step at which we consider applied inputs and determine the degree to which they belong in each of the appropriate fuzzy sets via membership functions. For example if we have an input value of 45 as height, then according to Figure 3, the results are 0.8 Short, 0.1 Medium and 0 tall.

**Step-2 Apply Rules:** "If–then" rules specify a relationship between the input and output for fuzzy sets. The "if" part of the rule, "x is A," is called the antecedent, while the "then" part of the rule, "y is B," is called the consequent or conclusion. . If a rule has more than one part, for example, "If x is A and y is B then z is C", the fuzzy logical operators are applied to evaluate the composite firing strength of the rule. The purpose of applying rules is to find out the degree to which the antecedent is satisfied for each rule.

**Step-3 Apply Implication Method:** The implication method is defined as the shaping of the output membership functions on the basis of the rule's firing strength. The input for the implication process is a single number given by the antecedent, and the output is a fuzzy set. Two commonly used methods of implication are the minimum and the product. Figure 3 shows the Mamdani Min-Max-Min Rule execution process.

**Step-4 Aggregate All Outputs:** Aggregation is a process whereby the outputs of each rule are unified. Aggregation occurs only once for each output variable. The input for the aggregation process is the truncated output fuzzy sets returned by the implication process for each rule. The output of the aggregation process is the combined output fuzzy set.

**Step-5 Defuzzify:** The input for the defuzzification process is a fuzzy set (the aggregated output fuzzy set), and the output of the defuzzification process is a value obtained by using a defuzzification method such as the centroid, height, or maximum.

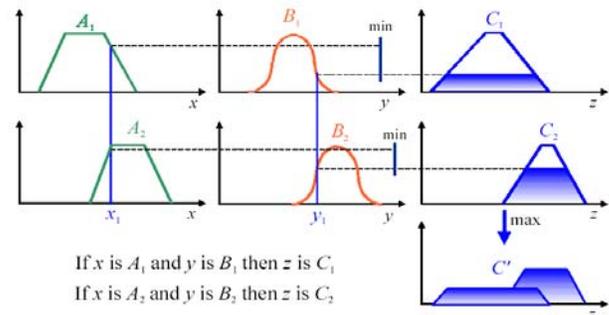

**Figure 3: Mamdani min-max-min rule execution process**

## 3. Software Product Line Decision Support Tool (SPLDST)

A fuzzy logic based tool intended to measure the performance of the software product line process is designed and developed on the basis of essential activities performed during software product line development. Since the software product line has three essential activities, core asset development, product development and management, then the three-dimensional approach for the process assessment of a software product line proposes to:

- Process assessment of individual activities like core asset development, product development and management.

- Process assessment of software product line as a function of the three-essential activities of core asset development, product development and management.

### 3.1 Software Product Line Process Input

Every fuzzy logic system requires certain input for processing; therefore, in order to take input in the form of quantitative data, specific questions are designed based on the fundamental activities performed during the development and management levels. The questions are divided into the three categories of core assets, product development and management, as presented in Table 1.

**Core Asset Development Input Questions:** These questions assess the general principles required to develop and manage a core asset repository in order to facilitate the reuse of core assets during software product development activity. The questions assess the components present in the core asset repository, specifically, their criteria for qualification.

**Product Development Input Questions:** These questions assess the qualification criteria for the products developed during software product line activity based on the core assets present in the core asset repository. They are used to evaluate the range of the products developed and the characteristics of the software product line.

**Management Input Questions:** These questions evaluate the essential management activities that must be followed and implemented to enable effective utilization of the software product line concept. They also assess the associated processes that are required for software product development and management.

### 3.2 Architecture of the SPLDST

The top-level block diagram of the SPLDST is shown in Figure 4. The overall software product line process assessment activity is divided into a set of three assessments activities of core asset, product development and management. The detailed architecture of the SPLDST is illustrated in Figure 5, showing the flow of information at each stage during the process assessment of a software product line project. All seventeen questions presented in Table 1 are applied to the two-variable fuzzy logic system, where intermediate outputs are collected and applied to the next stage. Individual activities such as core asset development, product development and management are evaluated and processed together to produce the final software product line process assessment. The structure of the two-variable fuzzy logic system is shown in Figure 6. The fuzzy logic system contains the fuzzy rule base, fuzzification, the fuzzy engine and the defuzzification sub-system. It requires two variables as input, which can be any combination of two questions presented in Table 1. Then, it performs a fuzzification process, which converts the input to a fuzzy membership mapping, which, in turn, is applied to the inference engine. The inference engine interacts with the rule base to select the applicable rules based on the input variable values; the fuzzy output is then defuzzified to retrieve the final output. Figure 7 illustrates the intermediate calculation steps of SPLDST.

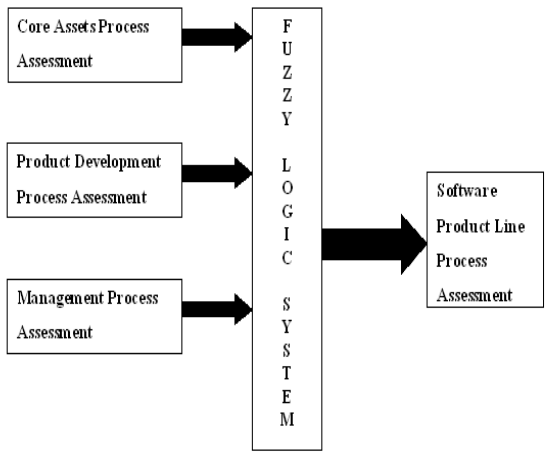

**Figure 4: Architecture of fuzzy logic based software product line decision support tool**

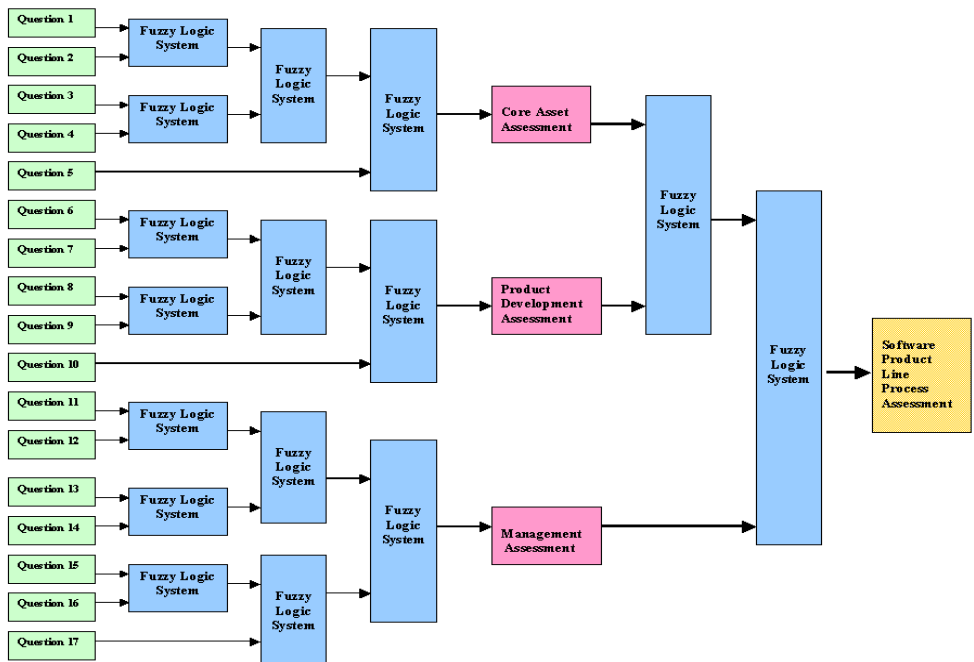

**Figure 5: Detailed architecture of SPLDST and structure of information flow**

**Table 1: Software Product Line Process Input Questions**

| | **Core Asset Development Input Questions** |
|---|---|
| 1. | Are all of the core assets within the software product line repository and are the resulting products consistent with the scope of the software product line? |
| 2. | Do all the components present in the core asset repository define the variability mechanism and tailor them for effective utilization? |
| 3. | Do all the COTS present or added into the core asset repository satisfy the cost-benefits ratio for the organization? |
| 4. | Is the core asset repository constantly updated with the addition of new assets as the product line progresses? |
| 5. | Does a version control management system keep track of the core asset development and utilization history? |
| | **Product Development Input Questions** |
| 6. | Do all the products within the software product line share a common architecture? |
| 7. | Does the variation among products remain within the scope of the software product line? |
| 8. | Is every product released from the product line an effective business decision for the organization? |
| 9. | Does the software product line produce a considerable number of products; in other words, do they produce more than one product? |
| 10. | Does every product released from the software product line meet the qualification criteria of the organization? |
| | **Management Input Questions** |
| 11. | Is there a configuration management system established to handle the configuration management issues present in the software product line? |
| 12. | Is a comprehensive description and analysis of the domain performed for the software product line? |
| 13. | Does the ROI (Return on Investment) of the software product line meet the organization's financial goal? |
| 14. | Are the requirements of the software product line clearly defined, analyzed, specified, verified and managed? |
| 15. | Does the requirement of the software product line define the fundamental products and their features within the product line? |
| 16. | Does the organizational structure support the software product line's concepts and principles? |
| 17. | Are the essential activities of software product line development performed iteratively? |

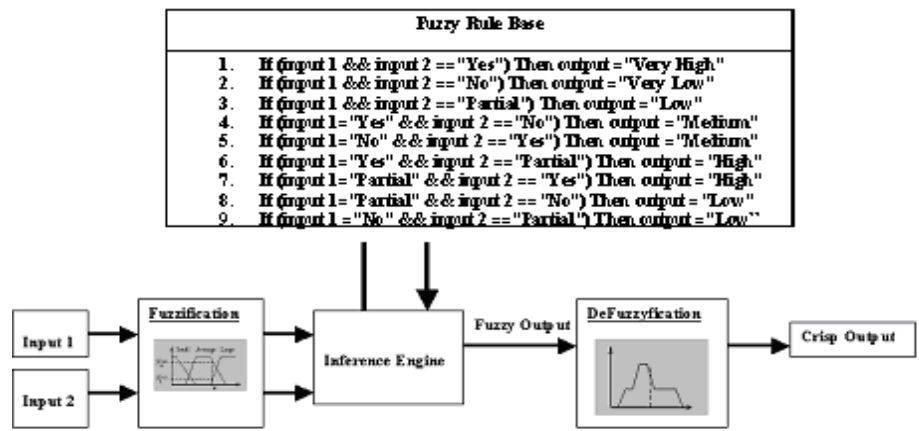

**Figure 6: Structure of fuzzy logic system block**

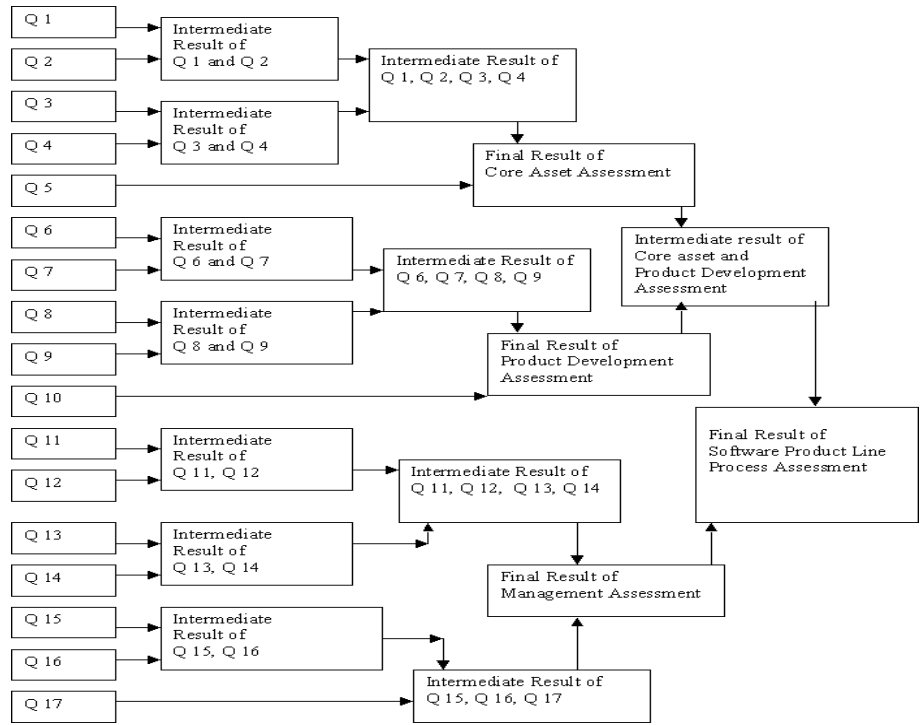

**Figure 7: SPLDST intermediate calculation steps**

## 3.3 Input Variable Mapping

The input for the fuzzy logic system depends on the values entered for each question, which ranges from 0 to 50. The value of 0 reflects the lowest ranking, whereas the value of 50 corresponds to the highest rating for a particular activity in the software product line process. The values in the range from 0 to 50 reflect the extent of a person's agreement with the questions relating to product line activities. As presented in Table-I, the input values are divided into three linguistic categories: "yes", "no" and "partial":

- **Yes** means that the activity is completely performed and is represented in the range of 33.0 to 50.0.

- **Partial** means that the activity is only partly performed and is represented in the range of 16.5 to 38.0.

- **No** means that the activity is not performed and is represented in the range of 0 to 21.5.

A trapezoid function is used to represent the mapping between the fuzzy membership, in the range of 0 to 1, and the input values, in the range of 0 to 50. Equation-I represents the mathematical model of the trapezoid function. The values of the variables a, b, c and d define the shape of the trapezoid. The graphical representation of the trapezoid function, along with variables a, b, c, and d, is shown in Figure 8, which illustrates that the choice of the variables a, b, c and d determines the shape of the trapezoid. Table 2 shows the distribution of the linguistic variables "yes", "no" and "partial" in the range of 0 to 50 and describes the values for the variables a, b, c and d in Equation-I, creating a mapping between linguistic variables and fuzzy membership values. Figure 9 illustrates the distribution of the input linguistic variables "yes", "no", and "partial" in the range of 0 to 50 and the fuzzy membership mapping in the range of 0 to 1.

$$f(x;a,b,c,d) = \begin{cases} 0 & \text{for } x < a \\ \frac{x-a}{b-a} & \text{for } a \leq x < b \\ 1 & \text{for } b \leq x < c \\ \frac{d-x}{d-c} & \text{for } c \leq x < d \\ 0 & \text{for } d \leq x \end{cases}$$

**Equation- I**

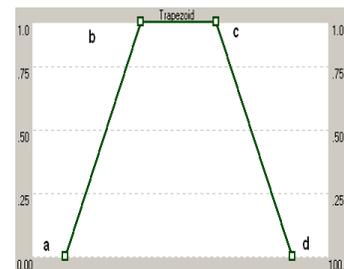

**Figure 8: Trapezoid function**

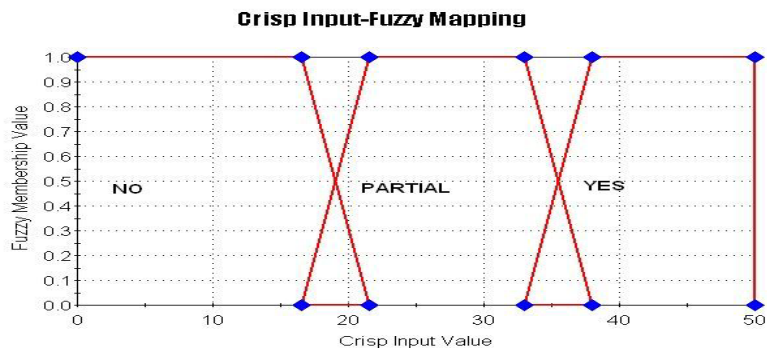

**Figure 9: Input fuzzy membership mapping**

**Table 2: Input Values Linguistic and Fuzzy Membership**

| Linguistic Value | Value Range | Trapezoid Function Variable Values For Input Fuzzy Membership Mapping | | | |
|---|---|---|---|---|---|
| | | a | b | C | D |
| No | 0 to 21.5 | 0.0 | 0.0 | 16.5 | 21.5 |
| Partial | 16.5 to 38.0 | 16.5 | 21.5 | 33.0 | 38.0 |
| Yes | 33.0 to 50 | 33.0 | 38.0 | 50.0 | 50.0 |

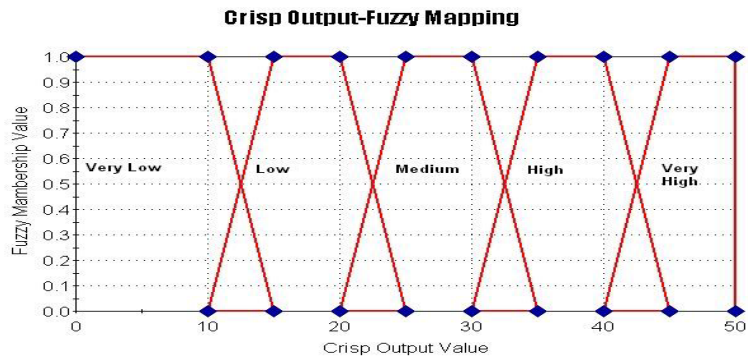

**Figure 10: Output fuzzy membership mapping**

**Table 3: Output Values Linguistic and Fuzzy Membership**

| Linguistic Value | Value Range | Trapezoid Function Variable Values For Output Fuzzy Membership Mapping | | | |
|---|---|---|---|---|---|
| | | a | b | C | D |
| Very Low | 0.0 to 15.0 | 0.0 | 0.0 | 10.0 | 15.0 |
| Low | 10.0 to 25.0 | 10.0 | 15.0 | 20.0 | 25.0 |
| Medium | 20.0 to 35.0 | 20.0 | 25.0 | 30.0 | 35.0 |
| High | 30.0 to 45.0 | 30.0 | 35.0 | 40.0 | 45.0 |
| Very High | 40.0 to 50.0 | 40.0 | 45.0 | 50.0 | 50.0 |

### 3.4 Output Variable Mapping

The output of the system is selected to fall in the range of 0 to 50. The output values are divided into five linguistic categories: very low, low, medium, high and very high. These values are in the range of 0 to 50, as described below:

- **Very Low:** defined in the interval of 0.0 to 15.0
- **Low:** defined in the interval of 10.0 to 25.0
- **Medium**: defined in the interval of 20.0 to 35.0
- **High:** defined in the interval of 30.0 to 45.0
- **Very High**: defined in the interval of 40.0 to 50.0

A trapezoid function is used to represent the mapping between the fuzzy membership, in the range of 0 to 1, and the output values, in the range of 0 to 50. Table 3 illustrates the distribution of the linguistic output variables "very low", "low", "medium", "high" and "very high" in the range of 0 to 50 and shows the values for variables a, b, c and d in Equation-I, creating a mapping between linguistic variables and fuzzy membership values. Figure 10 illustrates the distribution of output linguistic variables in the range of 0 to 50 and fuzzy membership mapping in the range of 0 to 1.

### 3.5 Fuzzy Rule Base

The fuzzy knowledge rule base contains fuzzy logic rules for the reasoning of the Software Product Line Decision Support Tool. The rules generally define a combination of the input pattern and the respective output. On the basis of the combination of input, appropriate output mapping is defined in the fuzzy logic rules. There are nine rules for the SPLDST, and it is important to note here that the "and" operator is used in the structure of the rule. The truth table of the rule base is as follows:

| Input 1 | Input 2 | Output |
|---------|---------|--------|
| Yes | Yes | Very High |
| No | No | Very Low |
| Partial | Partial | Low |
| Yes | No | Medium |
| No | Yes | Medium |
| Yes | Partial | High |
| Partial | Yes | High |
| Partial | No | Low |
| No | Partial | Low |

The terms "input 1" and "input 2" are used for the combined values of any two questions presented in Table 1, when applied to the input of the fuzzy logic system. Input_1 and input_2 fall in the linguistic domain of the terms "yes", "no" and "partial". On the other hand, the output falls in the linguistic domain of "very low", "low", "medium", "high" and "very high".

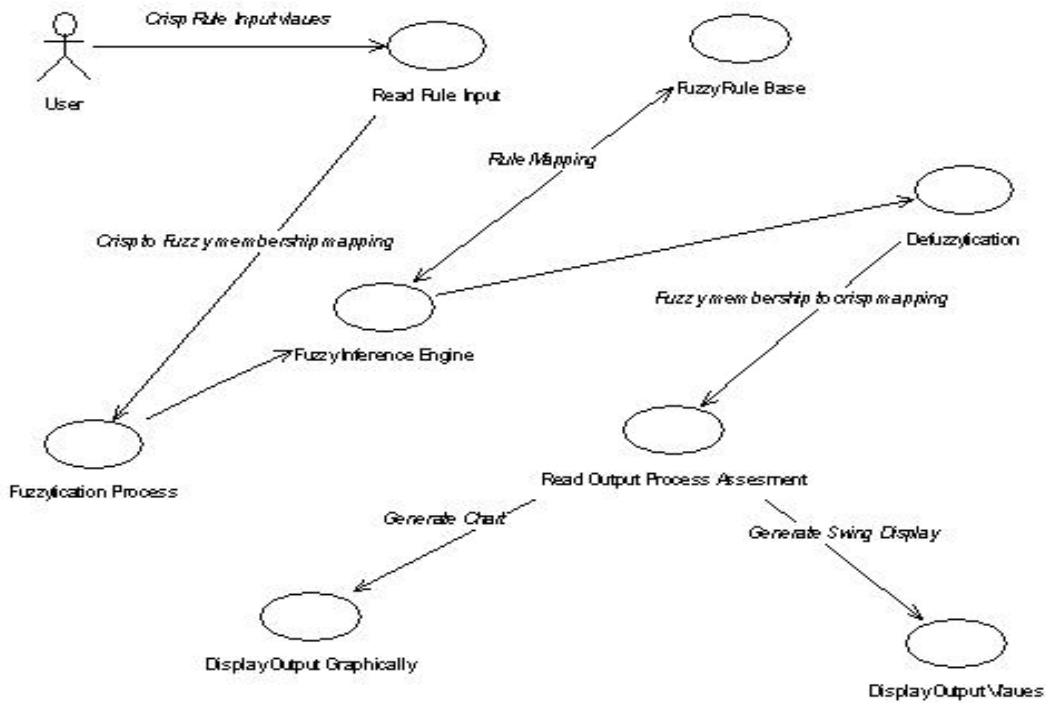

**Figure 11: Use case diagram of software product line decision support tool**

## 4. Visual Model of the SPLDST

The visual model of the SPLDST is represented in Unified Modelling Language (UML). The use case diagram of the SPLDST describes the system functionality as a set of various tasks that the system must perform, and it indicates the actors that interact with the system in order to complete the tasks. Each use case indicated on the diagram represents a single task that the system needs to carry out, such as rule input, the fuzzy rule base, fuzzification, the fuzzy inference engine, and defuzzification. Some use cases may include or continue a task represented by another use case. For example, in order to execute the fuzzy inference engine, the fuzzy rule information is required from the rule base. Figure 11 represents the top-level use case diagram of the Software Product Line Process Assessment Application. The sequence diagram of the SPLDST describes the sequence of actions that occur in the system. It also illustrates the order in which the requests for the procedures and the procedures themselves occur. The Sequence diagram of the SPLDST is shown in Figure 12, which illustrates the dynamic behaviour of the system. In Figure 12, the x-axis shows the life of the represented object and the y-axis depicts the sequence in which the objects were created. Figure 13 illustrates the class diagram of SPLDST, which gives vital information on which classes are created and how the classes interact with each other in the form of message passing and inheritance.

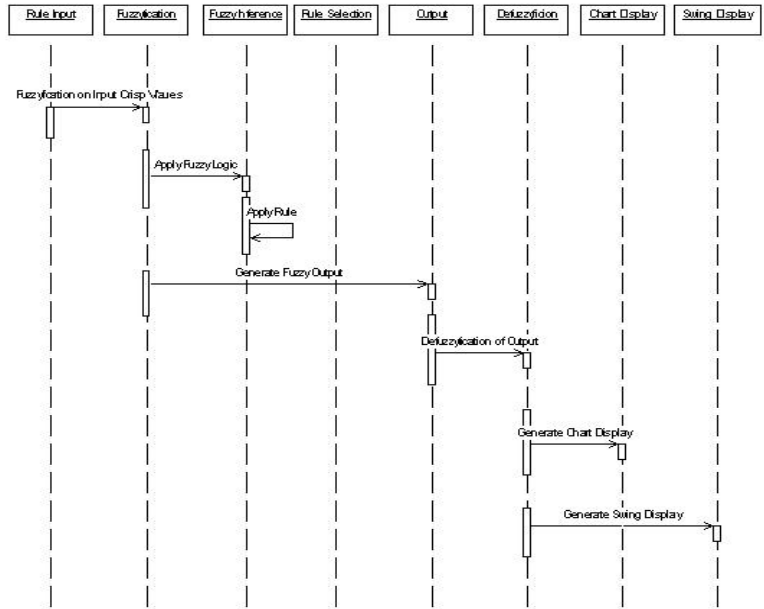

**Figure 12: Sequence diagram of software product line decision support tool**

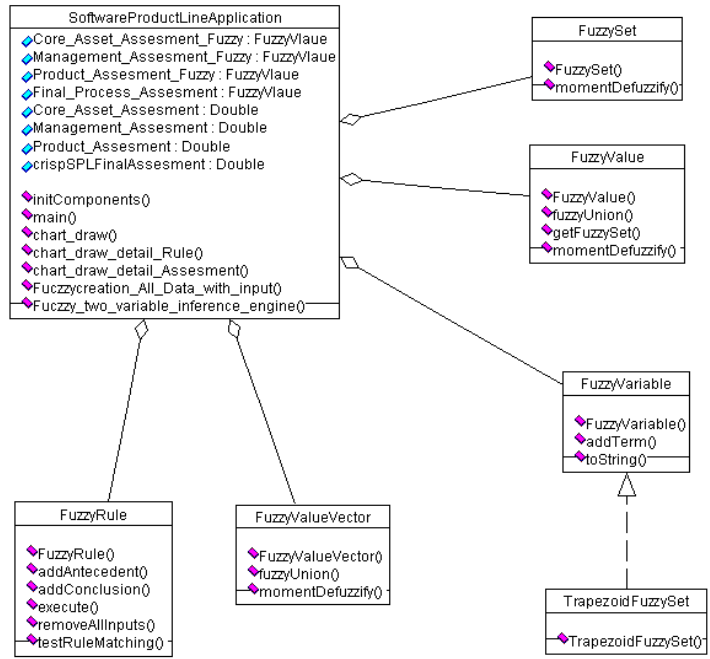

**Figure 13: Class diagram of software product line decision support tool**

## 4.1 SPLDST Programming Description

The NRC FuzzyJ [20] toolkit developed by National Research Council of Canada Institute of Information Technology is a set of Java classes to support the capability of handling fuzzy concepts and reasoning in Java application environment. The goal of the toolkit is to provide a useful tool for exploring the ideas of fuzzy logic and fuzzy reasoning in a Java development environment. The various classes present in the toolkit provide the basic foundation for developing fuzzy logic applications in Java. The toolkit is used in software product line decision support tool to handle and implement fuzzy logic concepts.

### 4.1.1 Fuzzy Input Variable

An input variable is created by using nrc.fuzzy.FuzzyVariable. The input variable further constructs three linguistic input variables, i.e. "No", "Partial" and "Yes". The steps involved in creating a fuzzy variable with associated linguistic expression are as follows:

Create an object of *nrc.fuzzy.FuzzyVariable* and define the range of crisp value.

```
FuzzyVariable Input_1 = new FuzzyVariable("Input_1", 0.0, 50.0)
```

The whole width of 0 to 50 of fuzzy variable is divided into three linguistic variables. A trapezoid function is used to define mapping between crisp value and fuzzy membership. The *addTerm()* method is applied to the *nrc.fuzzy.FuzzyVariable* object to create three linguistic variables. The constructor method of *nrc.fuzzy.TrapezoidFuzzySet* class is used to define range for crisp values to linguistic variables.

```
Input_1.addTerm ("No", new TrapezoidFuzzySet(0.0, 0.0,16.5, 21.5));
Input_1.addTerm("Partial",new TrapezoidFuzzySet(16.5, 21.5,33.0, 38.0));
Input_1.addTerm("Yes", new TrapezoidFuzzySet (33.0, 38.0,50.0,50.0));
```

### 4.1.2 Fuzzy Output Variable

A fuzzy variable is created to represent output. Five linguistic variables, i.e., "Very Low", "Low", "Medium", "High" and "Very High" are mapped and distributed over the range of 0 to 50. Similar to the input variable, the mapping between crisp output value and fuzzy membership is performed and all the linguistic variables maintain a value of fuzzy membership 1 for a certain interval. The following code segment shows how the output variable is created and how output linguistic variables are added and distributed over the range of 0 to 50.

```
FuzzyVariable output = new FuzzyVariable("output", 0.0, 50.0);

output.addTerm("Very Low", new TrapezoidFuzzySet(0.0, 0.0,10.0, 15.0));
output.addTerm("Low", new TrapezoidFuzzySet(10.0, 15.0,20.0, 25.0));
output.addTerm("Medium", new TrapezoidFuzzySet(20.0, 25.0,30.0, 35.0));
output.addTerm("High", new TrapezoidFuzzySet(30.0, 35.0,40.0, 45.0));
output.addTerm("Very High", new TrapezoidFuzzySet (40.0, 45.0,50.0,50.0));
```

### 4.1.3 Fuzzy Rule Base Implementation

Fuzzy rules are created with the help of *nrc.Fuzzy.FuzzyRule* class and its two associated methods, i.e. *addAntecedent()* and *addConclusion()*. A total of nine rules are created to map two inputs into five output linguistic variable as shown below:

```
FuzzyRule Fuzzy1=new FuzzyRule();
FuzzyRule Fuzzy2=new FuzzyRule();
FuzzyRule Fuzzy3=new FuzzyRule();
FuzzyRule Fuzzy4=new FuzzyRule();
FuzzyRule Fuzzy5=new FuzzyRule();
FuzzyRule Fuzzy6=new FuzzyRule();
FuzzyRule Fuzzy7=new FuzzyRule();
FuzzyRule Fuzzy8=new FuzzyRule();
FuzzyRule Fuzzy9=new FuzzyRule();

Fuzzy1.addAntecedent(new FuzzyValue(Input_1,"Yes"));
Fuzzy1.addAntecedent(new FuzzyValue(Input_2,"Yes"));
Fuzzy1.addConclusion(new FuzzyValue(output,"Ver High"));

Fuzzy2.addAntecedent(new FuzzyValue(Input_1,"No"));
Fuzzy2.addAntecedent(new FuzzyValue(Input_2,"No"));
Fuzzy2.addConclusion(new FuzzyValue(output,"Very Low"));

Fuzzy3.addAntecedent(new FuzzyValue(Input_1,"Partial"));
Fuzzy3.addAntecedent(new FuzzyValue(Input_2,"Partial"));
Fuzzy3.addConclusion(new FuzzyValue(output,"Low"));

Fuzzy4.addAntecedent(new FuzzyValue(Input_1,"Yes"));
Fuzzy4.addAntecedent(new FuzzyValue(Input_2,"No"));
Fuzzy4.addConclusion(new FuzzyValue(output,"Medium"));

Fuzzy5.addAntecedent(new FuzzyValue(Input_1,"No"));
Fuzzy5.addAntecedent(new FuzzyValue(Input_2,"Yes"));
Fuzzy5.addConclusion(new FuzzyValue(output,"Medium"));

Fuzzy6.addAntecedent(new FuzzyValue(Input_1,"Yes"));
Fuzzy6.addAntecedent(new FuzzyValue(Input_2,"Partial"));
Fuzzy6.addConclusion(new FuzzyValue(output,"High"));

Fuzzy7.addAntecedent(new FuzzyValue(Input_1,"Partial"));
Fuzzy7.addAntecedent(new FuzzyValue(Input_2,"Yes"));
Fuzzy7.addConclusion(new FuzzyValue(output,"High"));

Fuzzy8.addAntecedent(new FuzzyValue(Input_1,"Partial"));
Fuzzy8.addAntecedent(new FuzzyValue(Input_2,"No"));
Fuzzy8.addConclusion(new FuzzyValue(output,"Low"));

Fuzzy9.addAntecedent(new FuzzyValue(Input_1,"No"));
Fuzzy9.addAntecedent(new FuzzyValue(Input_2,"Partial"));
Fuzzy9.addConclusion(new FuzzyValue(output,"Low"));
```

### 4.1.4 Fuzzy Inference Engine Implementation

The fuzzy inference engine of SPLDST is implemented using *FuzzyRuleExecutor* interface, which provides an "execute" method that accepts a *FuzzyRule* object and returns a *FuzzyValueVector* composed of the actual conclusion of *FuzzyValues* for the rule using the Mamdani Min inference operator and Max-Min composition. It cycles the nine rules one by one and determines whether or not the rule is applicable. If the rule is applicable then an intermediate output is calculated by applying a Mamdani Min inference operator, which compares the two inputs and places their intersection as an output. Once all the rules have been applied, the collection of output values are applied through the Max-Min composition principle of Mamdani, which employs a centroid method to evaluate the final output.

```
Fuzzy1.removeAllInputs();
Fuzzy1.addInput(new FuzzyValue(Rule_1,new SingletonFuzzySet(input_1)));
Fuzzy1.addInput(new FuzzyValue(Rule_2,new SingletonFuzzySet(input_2)));

if(Fuzzy1.testRuleMatching())
{
FuzzyValueVector fvv = Fuzzy1.execute(new MamdaniMinMaxMinRuleExecutor());
Local_Assesment_of_inference_engine = fvv.fuzzyValueAt(0);
if(Final_Process_Assesment_of_inference_engine==null)
Final_Process_Assesment_of_inference_engine=Local_Assesment_of_inference_engine;
else
Final_Process_Assesment_of_inference_engine=Final_Process_Assesment_of_inference_engine.fuzzyUnion(Local_Assesment_of_inference_engine);
}
```

## 5. Case Studies and Tool Application

Four experiments were conducted in order to validate the results achieved from the SPLDST. The input questions, shown in Table 1, were distributed to a number of organizations to obtain data about the current process status within the organization. Some large and well-known organizations extensively involved in software development provided information with the mutual agreement of keeping the name of the organization confidential. For experimental purposes, the organizations are coded as "A", "B", "C" and "D". We asked the respondents to consult major sources of data in their organization, such as documents, plans, models, and actors, before responding to a particular question in order to reduce the tendency to overestimate or underestimate when filling in questionnaires. In case where we received multiple responses within one organization, we used an average of all the responses received from that particular organization.

### 5.1 Case Study –I

Case study-I was conducted on the data received from Organization "A", which is a famous company in electrical engineering and has been involved in the development of electronic equipment for a long time. Their expertise in computer controlled technology allowed them to establish software development centres, where large numbers of people are currently involved in producing software for their customized equipment. Overall, the SPLDST has evaluated the company's maturity process as low. More specifically, the results in Table 5 indicate that core asset development activity is performed at a maturity level between medium and high, meaning that the medium level has been achieved and the high level is close to being achieved. The product development activity is performed at the maturity level of medium, and the management activity has a maturity level of very Low, which lowers the overall process assessment to low. Figure 15 illustrates the output results of Case Study – I. The maturity level of the core assets, the product development, and the management activities are plotted along with the overall maturity assessment of the software product line. The main conclusion of Case Study – I indicates that Organization "A" can improve the overall software product line process by concentrating more on management activity. Figure 16 illustrates the internal processing sequence of the SPLDST, in which a combination of two questions from Table 1 is placed in the input of the two-variable fuzzy logic system described in Figure 6. The intermediate outputs are collected and given to the two-variable fuzzy logic system at the next level. This procedure continues until we collect the individual software product line activity, such as core asset development, management and product development assessment. These activities are later applied to the fuzzy logic system in order to obtain the overall software product line process assessment.

### 5.2 Case Study – II

The data received from Organization "B" is used in Case Study-II to assess its software product line process. Organization "B" is a highly significant company in the communications industry and has been involved in the development of communication related equipment for many years. They have established in-house software development centres where large numbers of people are currently involved in producing software for their customized communication equipment. Overall, the SPLDST evaluated the process maturity of Organization "B" as very high. More specifically, the results in Table 6 show that both core asset and product development activity is performed at the maturity level of high. The management activity is performed at a maturity level between high and very high, meaning that as a whole, the company has achieved the level of high and they are close to the level of very high.

**Table 4: Software Product Line Process Input for Case Studies**

| Rule # | Case Study-I | Case Study-II | Case Study-III | Case Study-IV |
|---|---|---|---|---|
| 1 | 35 | 40 | 32.5 | 40 |
| 2 | 40 | 40 | 27.5 | 30 |
| 3 | 25 | 15 | 30 | 35 |
| 4 | 35 | 30 | 37.5 | 30 |
| 5 | 25 | 50 | 40 | 20 |
| 6 | 40 | 15 | 37.5 | 40 |
| 7 | 10 | 15 | 32.5 | 35 |
| 8 | 5 | 30 | 30 | 35 |
| 9 | 50 | 50 | 35 | 30 |
| 10 | 45 | 40 | 37.5 | 30 |
| 11 | 30 | 50 | 32.5 | 25 |
| 12 | 10 | 40 | 35 | 20 |
| 13 | 15 | 40 | 30 | 30 |
| 14 | 20 | 30 | 35 | 35 |
| 15 | 30 | 40 | 32.5 | 35 |
| 16 | 35 | 45 | 30 | 35 |
| 17 | 7 | 25 | 37.5 | 35 |

**Table 5: Results of Software Product Line Process Assessment of Case Study-I**

| Activity | Result | Linguistic Output |
|---|---|---|
| Core Asset Process Assessment | 34.84 | Medium to High |
| Product Development Process Assessment | 29.72 | Medium |
| Management Process Assessment | 8.64 | Very Low |
| Software Product Line Process Assessment | 17.5 | Low |

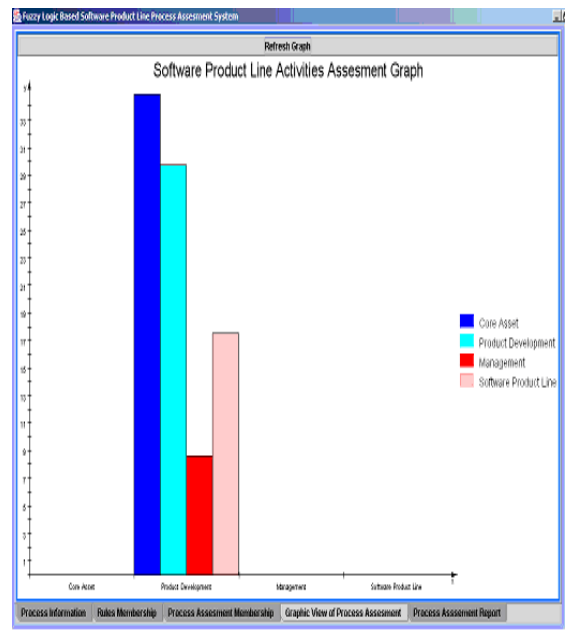

**Figure 14: Data entry input screen of SPLDST for case study-1**

**Figure 15: Graphical output screen of SPLDST for case study-1**

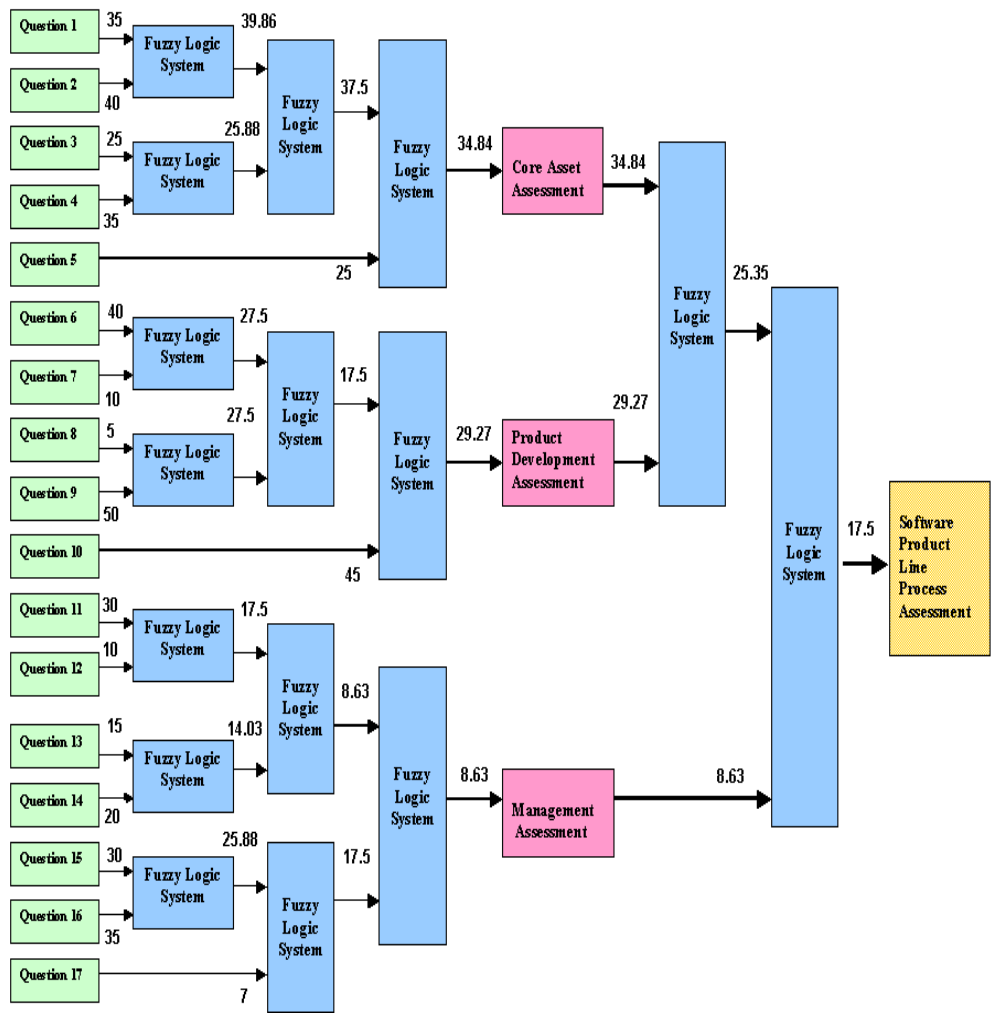

**Figure 16: Case study-1 processing sequence and intermediate results**

### 5.3 Case Study – III

Organization "C" is a worldwide software development and hardware manufacturing organization that has been involved in software and hardware development for decades. They have several software development centres throughout the world and are actively involved in customized and commercial software. The SPLDST evaluated the process maturity of organization "C" at the medium level. The results presented in Table 7 indicate that core asset development activity is performed at a maturity level of high. Product development activity is also performed at a maturity level of medium to high, meaning that the medium level has been achieved and the high is close to being achieved. However, the management activity is performed at the maturity level of low. Thus, Case Study – III indicates that Organization "C" has achieved a medium level of process maturity, demonstrating that there is a need to improve the management activity in order to increase the overall maturity level of the organization.

### 5.4 Case Study – IV

Organization "D" provided input values for the SPLDST to conduct Case Study-IV. Organization "D" is a pharmaceutical distribution centre with their business based mainly on E-commerce technology. They have developed multiple E-commerce web sites based on a common architecture with minor variability in requirements. Table 8 indicates that core asset development activity is performed at a maturity level of Medium. Also, the product development activity is performed at a higher maturity level ranging between medium and high, meaning that the medium level has been achieved and high level high has nearly been accomplished. However, the management activity is performed at the maturity level of low. The conclusion of the Case Study – IV illustrates that Organization "D" has achieved a low maturity level, indicating a need to improve the management and core asset development activity in order to increase the overall maturity level of the organization.

Table 7: Results of Process Assessment of Case Study – III

| Activity | Result | Linguistic Output |
|---|---|---|
| Core Asset Process Assessment | 37.5 | High |
| Product Development Process Assessment | 34.84 | Medium to High |
| Management Process Assessment | 17.5 | Low |
| Software Product Line Process Assessment | 27.07 | Medium |

Table 8: Results of Process Assessment of Case Study – IV

| Activity | Result | Linguistic Output |
|---|---|---|
| Core Asset Process Assessment | 25.65 | Medium |
| Product Development Process Assessment | 34.84 | Medium to High |
| Management Process Assessment | 17.5 | Low |
| Software Product Line Process Assessment | 17.5 | Low |

Table 9: SPLDST Process Assessment Results

| Organization | Software Product Line Process Assessment Level |
|---|---|
| "A" | 2 (Low) |
| "B" | 5 (Very High) |
| "C" | 3 (Medium) |
| "D" | 2 (Low) |

## 6. Validity Analysis

The two most important aspects of precision in the questionnaire-based process assessment approaches are reliability and validity. Reliability refers to the ability to reproduce a measurement, whereas validity refers to the agreement between the value of a measurement and its true value. The reliability of the questionnaire specifically designed for this study was evaluated by using the approach of internal consistency analysis. Internal consistency analysis was performed using the coefficient alpha [15], which was measured as 0.14. Since the value of the coefficient alpha ranges from 0 to 1, 0.14 is a relatively low yet positive measure of validity. The potential cause for this relatively low value has to do with our small sample size.

Construct validity, according to Campbell and Fiske [16], occurs when the scale items in a given construct are the same direction (for reflective measures) and, thus, highly correlate. Principal component analysis [17] was performed in order to observe a measure of convergent validity. We used eigen values [19] and scree plots [18] as reference points to observe the construct validity. According to Kaiser Criterion [19], any component having an eigen value greater then one is retained. Eigen value analysis revealed that three factors have eigen values of 1.69, 1.08 and 1.0, and the scree plots clearly show a truncation at the third component. In our questionnaire, we have three major factors of core assets, product development and management; therefore, the construct validity analysis supports the structure of the questionnaire. Thus, the convergent validity can be regarded as sufficient.

## 7. Conclusion and Future Work

In this work, we presented a decision support tool for the software product line process in order to evaluate the current process maturity level within an organization. The SPLDST provides a direct mechanism to assess the process maturity level of a software product line. Table 9 shows the conclusion of the experiments conducted by using the SPLDST on industrial data from well-known software development organizations. The maturity levels of "very low", "low", "medium", "high" and "very high" depict an organization's ability to successfully adopt a software product line process. This research will enable an organization to understand the effectiveness of the development process and allow them to predict the outcome of establishing and maintaining a software product line. Furthermore, it will enable a company to discover and monitor the strengths and weaknesses of various activities performed during software product line development and help them to improve the productivity of the development process. Currently, we are working on developing a comprehensive process maturity model, specifically for the process assessment of a software product line. The aim of this research is to identify the certain process areas of the software product line along with the specific and general practices carried out in each area in order to collect the process data for assessment.

## Authors

Faheem Ahmed received his MS (2004) and Ph.D. (2006) in Electrical Engineering from the University of Western Ontario, London, Canada. Currently he is working at College of Information Technology, UAE University, United Arab Emirates as assistant professor. Ahmed had many years of industrial experience holding various technical positions in software development organizations. During his professional career he has been actively involved in the life cycle of software development process including requirements management, system analysis and design, software development, testing, delivery and maintenance. Dr. Ahmed has authored and co-authored many peer-reviewed research articles in leading journals and conference proceedings in the area of software engineering. His current research interests are Software Product Line, Software Process Modeling, Software Process Assessment, and Empirical Software Engineering. He is a member of the IEEE.

Luiz Fernando Capretz has extensive experience in the engineering of software. He worked in Brazil, Argentina, England, Japan, and Canada as a practitioner, manager, and educator in the field of software engineering. Dr. Capretz has a B.Sc. from UNICAMP – Brazil, Masters from INPE – Brazil, and Ph.D. from the University of Newcastle – U.K.; all degrees in computer science. Dr. Capretz is the Director of the Software Engineering Program at the University of Western Ontario (Canada), and is currently on sabbatical leave as a visiting associate professor at the University of Sharjah (United Arab Emirates). He is a senior member of the IEEE, member of the ACM, MBTI certified practitioner, and licensed P.Eng. in Canada.